\documentclass[manuscript]{aastex}
\shorttitle{Barnard's Loop and the WIM}
\shortauthors{O'Dell et~al.}
\begin{document}
\title
{Physical Conditions in Barnard's Loop, Components of the Orion-Eridanus Bubble, and Implications for the WIM Component of the ISM
\thanks{
Based in part on observations obtained at the Cerro Tololo Inter-American Observatory, which is operated by the Association of Universities for Research in Astronomy, Inc., under a Cooperative Agreement with the National Science Foundation.}}

\author{C.~R.~O'Dell}
\affil{Department of Physics and Astronomy, Vanderbilt University, Nashville, TN 37235}
\author{G.~J.~Ferland}
\affil{Department of Physics and Astronomy, University of Kentucky, Lexington, KY 40506}
\author{R.~L.~Porter}
\affil{265 North Harris Street, Athens, GA 30601}
\and
\author{P.~A.~M.~van~Hoof}
\affil{Royal Observatory of Belgium, Ringlaan 3, 1180 Brussels, Belgium}


\newcommand{\kms}{\ensuremath{\mathrm{km\ s}^{-1}}}
\newcommand{\pcc}{\ensuremath{\mathrm{cm}^{-3}}}
\newcommand\Halpha{H$\alpha$}
\newcommand\Hbeta{H$\beta$}
\newcommand\hanii{H$\alpha$+[N~II]}
\newcommand\sbunits{$\rm photons~cm ^{-2}~s^{-1}~sr^{-1}$}
\newcommand\funits{$\rm erg~cm^{-2}~s^{-1}$}
\newcommand\cmq{$\rm cm^{-3}$}
\newcommand\cmsq{$\rm cm^{-2}$}
\newcommand\ori{$\theta ^{1}$~Ori~C}
\newcommand\oriA{$\theta ^{2}$Ori~A}
\newcommand\emunits{$\rm cm^{-6}~pc$}
\newcommand\pers{$\rm s^{-1}$}
\newcommand\bub{Orion-Eridanus Bubble}
\newcommand\Tstar{T$\rm _{star}$}
\newcommand\Te{T$\rm _{e}$}
\newcommand\hour{\ensuremath{^\mathrm{h}}}
\newcommand\minute{\ensuremath{^\mathrm{m}}}
\newcommand\second{\ensuremath{^\mathrm{s}}}
\newcommand\sbu{photons $\rm cm^{-2}~s^{-1}~steradian^{-1}$}

\begin{abstract}
We have supplemented existing spectra of Barnard's Loop with high accuracy spectrophotometry of one new position. Cloudy photoionization models were calculated for a variety of ionization parameters and stellar temperatures and compared with the observations. After testing the procedure with recent observations of M43, we establish that Barnard's Loop is photoionized by four candidate ionizing stars, but agreement between the models and observations is only possible if Barnard's Loop is enhanced in heavy elements by about a factor of 1.4.

Barnard's Loop is very similar in properties to the brightest components of the Orion-Eridanus Bubble and the Warm Ionized Medium (WIM). We are able to establish models that bound the range populated in low-ionization color-color diagrams (I([SII])/I(\Halpha) versus I([NII])/I(\Halpha)) using only a
limited range of ionization parameters and stellar temperatures.

 Previously established variations in the relative abundance of heavy elements render uncertain
 the most common method of determining electron temperatures for components of the Orion-Eridanus Bubble and the WIM based on only the I([NII])/I(\Halpha) ratio, although we confirm that the lowest surface brightness components of the WIM are on average of higher electron temperature.

The electron temperatures for a few high surface brightness WIM components determined by direct methods are comparable to those of classical bright H~II regions. In contrast, the low surface brightness HII regions studied by the Wisconsin \Halpha\ Mapper are of lower temperatures than the classical bright HII regions.

\end{abstract}


\section{Introduction}

The object commonly known as Barnard's Loop was discovered photographically more than a century ago, originally by W. H. Pickering in 1889 \citep{she95} and then by E. E.~\citet{bar94}.  Although a popular object for imaging by amateur astronomers, its low surface brightness has delayed the definitive observations necessary for  developing a good model of its characteristics. A representative image is shown in Figure 1.  It is an arc of about 14\degr\  facing east and at a distance of 440 pc \citep{oh08} is 110 pc in length. The northern boundary is at a Galactic Latitude of -13\degr\ and lines of galactic longitude are at a position angle of about 62\degr. These facts mean that its two sides (north and south) extend over a distance of about 60 pc and the conditions of the local ambient interstellar medium will have changed significantly. The north edge is about twice as bright as the south, indicating a higher density there.  

The structure of Barnard's Loop was the subject of a study of the brightness distribution of ultraviolet light measured in a photograph obtained during the Gemini 11 manned space-flight. In this investigation \citet{oyh67} derived the distribution of interstellar dust under the assumption that the ultraviolet continuum is due to scattered light originating in the OB stars in the Orion constellation Belt and Sword regions. It was found that the dust density increases approximately as the square of the distance from the center of Barnard's Loop. Under the assumption of a typical gas to dust ratio \citet{oyh67} calculated that an ionization boundary should occur where one observes Barnard's loop to be brightest in \Halpha. It is expected that there would be neutral hydrogen
outside of this ionization boundary, which is consistent with older \citep{men58} and more recent \citep{hb97} studies of HI 21-cm emission from this region.  Optical emission line splitting of about 24 \kms\ has been determined \citep{mad06}, consistent with the interpretation of \citet{oyh67} that Barnard's Loop is a result of the compression of ambient  interstellar material pushed outward by radiation pressure force acting on the dust component, with the light originating from Sword and Belt region stars.  

Barnard's Loop is now understood to be only the bright eastern portion of a much larger irregular shell of material of 41\degr\ x 27\degr\  (380x220 pc) oriented east-west and thought \citep{ro79} to be expanding at between 15 and 23 \kms.
The entire object is called the Orion-Eridanus Bubble. The well delineated western part of the \bub\ shows the ion distribution of an ionization front \citep{mad06} surrounded by neutral material \citep{hb97}. The low velocity of expansion means that one would not expect mass-motion collisional excitation and ionization to be important, so that photoionization processes should dominate, but, the low velocities also make it difficult to discriminate between the radiation pressure driven model invoked by \citet{oyh67} and a mass-loaded shell of a supernova  remnant, the interpretation most commonly applied to explain the expansion \citep{mad06}.

The conditions of the gas in the \bub, as indicated by emission line ratios, appears to resemble \citep{mad06} those of the very low-density, hot component of the interstellar medium called the WIM (Warm Ionized Medium) and contrast with those in the bright Galactic H~II regions.  Therefore, explanation of the conditions within the \bub\ may lead to understanding  the heating processes within the WIM, which are presently not adequately understood \citep{haf09}. 

In this paper we will  (\S\ 2) summarize the results from other studies of Barnard's Loop and present new spectrophotometric observations, then use these observations to test (\S\ 3) models for the photoionization of the \bub, demonstrating that the 
nebular lines, except for [O~III], can be explained by photoionization by the best four candidate ionizing stars only if Barnard's Loop is enhanced in heavy elements by about a factor of 1.4.   In \S\ 4 we argue that the low surface brightness ubiquitous component of the interstellar medium (ISM) commonly known as the WIM is subject to the same physical processes as the Barnard's Loop, that the electron temperatures in WIM components are similar to those of the classical bright H~II regions, 
and that the common method of determining exact electron temperatures (\Te ) of WIM components is rendered uncertain because of variations in heavy element abundances comparable to those seen in well studied H~II regions.

\section{Observations}\label{obs}
Due to its low surface brightness Barnard's Loop has not been extensively studied with adequate spectral resolution. It is a popular object for imaging with wide-field cameras but usually with a filter that passes both the \Halpha\ emission line at 6563 \AA\  and the adjacent [N~II] doublet at 6548 \AA\ and 6583 \AA, making it impossible to develop a quantitative analysis. The fainter portions of the \bub\ have been even more difficult to study. In this section we will summarize previous observations of the region and present new material on Barnard's Loop.

\subsection{Spectrophotometry of Barnard's Loop}
\subsubsection{Previous Observations}
There have been two earlier studies that are particularly useful. The older study was at low spectral and high spatial resolution and the more recent at quite high spectral resolution but low spatial resolution. 

A photoelectric scanner study of several faint regions of the sky \citep{pei75} included useful results for one position in Barnard's Loop (shown in Figure 1) with two 5\farcs2 x 77\farcs6 slits separated by 168\arcsec. The spectral resolution is not stated but was adequate for separating the \Halpha\ and 6583 \AA\ lines. Although of apparently low signal to noise ratio (only the brightest lines were detected), this study extended to the bright [O~II] doublet at 3727 \AA. The average surface brightness of these two samples was $1.1\times 10^{7}$ \sbu\  in \Halpha\ or 140 Rayleighs (the latter units are often employed in studies of extended emission and are  4$\pi$/10$^{6}$ times the surface brightness expressed in photon rate).  The line flux ratios are presented in Table 1.

A scanning Fabry-Perot spectrophotometer (the Wisconsin \Halpha\ Mapper, known as WHAM) study was conducted by \citet{mad06} with a 1\degr\ diameter field and a spectral resolution of 12  \kms, measuring the [O~III] 5007 \AA,
HeI 5876 \AA, \Halpha\ 6563 \AA, [N~II] 6583 \AA, and [S~II] 6716 \AA\ lines. The \Halpha\ surface brightness of position 1 was 86.3$\pm$0.1 Rayleighs and for position 2 it was 228.1$\pm$0.2 Rayleighs. The relative fluxes are given in Table 1.

\subsubsection{New Spectrophotometry}
New spectroscopic observations were made with the Cerro Tololo Interamerican Observatory 1.5 m-telescope operated in cooperation with the SMARTS consortium.
The instrument was the Boller and Chivens spectrograph. Observations were made on 2008
November 24 with Grating G58 using the Loral 1K CCD detector. A GG395 glass filter was used to prevent second order flux from contaminating the first order flux that was targeted.
One pixel projected 1\farcs3 along the 429\arcsec\ long slit, while the slit width was 2\farcs15. The measured full width at half maximum intensity of the nebular lines were 6 \AA. The CTIO spectrophotometric standard star Feige 15 was observed nine times and the results were used to calibrate the nebular observations into energy units. The positions of the slit setting is shown in Figure 1 in addition to the position used for determining the background sky brightness. The sample was 6\fdg07 from the Trapezium stars that contain \ori, the earliest spectral type star in the Orion Belt and Sword region.

In order to facilitate sky subtraction and cosmic ray cleaning double exposures were made in the pattern of two 1800 s Barnard's Loop exposures followed by two 900 s sky exposures, a pattern repeated three times. Since it was considered most important to produce good sky subtractions near the \Halpha\ line, we scaled the sky observations to null out the OH band sky signal close to \Halpha\ prior to making the sky subtraction. This meant that other strong night sky lines that varied in brightness relative to the night sky OH emission were inadequately subtracted from the nebular spectra.

Data reduction was done using standard IRAF procedures\footnote{IRAF is distributed by the National Optical Astronomy Observatories, which is operated by the Association of Universities for Research in Astronomy, Inc.\ under cooperative agreement with the National Science foundation.}. The results of
these steps were calibrated spectra expressed in ergs \cmsq\ s$^{-1}$ pixel$^{-1}$\ that were then converted to surface brightness units and averaged over the entire length of the entrance slit. Only the 
H$\gamma$ 4340 \AA,  \Hbeta\ 4861 \AA, \Halpha\ 6563 \AA,  [N~II] 6583 \AA,  [N~II] 6548 \AA, [S~II] 6716 \AA, and [S~II] 6731 \AA\ lines were bright enough to be measured using the task "splot", which required de-blending the lines near \Halpha\ and the [S~II] doublet using task "deblend". The results from two spectra of the central Orion Nebula obtained during the same observing run were used to compare the derived surface brightness in the \Hbeta\ line with those obtained from the spectrophotometric study of \citet{b91} and the calibration of the Hubble Space Telescope WFPC2 emission line filters \citet{od99}, which use the \citet{b91} results as a standard. There was good agreement. The surface brightness in \Hbeta\ was  $5.3\times 10^{6}$ \sbu\ or 66.6 Rayleighs. The final averaged spectrum is shown in Figure 2. The negative values at 5577 \AA, 6300 \AA, and 6363 \AA\ reflect the uncertainties in subtracting the sky background as these [O~I] sky lines varied considerably in surface brightness and the sky signal at 5577 \AA\ was comparable to the nebula's signal at \Halpha\ 6563 \AA.  An upper limit for the nebula's [O~III] 5007 \AA\  and He~I 5876 \AA\ lines was determined by scaling the \Hbeta\ line and inserting it at the locations of those lines. The flux ratio of the 6716 \AA\ and 6731 \AA\ lines was within 1\%\ of the theoretical low density limit of 1.50. The relative line intensities, normalized to \Hbeta\ are given in Table 1.

\begin{table*}
 \centering
 \begin{minipage}{126mm}
\caption{Flux Ratios for the Barnard's Loop Samples.}
\label{tableone}
 \begin{tabular}{@{}lccccc@{}}
 \hline
Wavelength (\AA) & Source & This Study & Peimbert et~al. 1975 & WHAM1 & WHAM2\\
\hline
3727 & [O~II]              & ----                & 0.60    & ----   & ----\\
4340 & H$\gamma$ & 0.51               & 0.51   & ----   & ----\\
4861 & \Hbeta            &  1.00           & 1.00      & ----   & ----\\
5007 & [O~III]              & $<$0.007 & $\la$0.22 & 0.020 & 0.004\\
5876 & HeI                  & $<$0.007 & ----    & 0.015 & 0.004\\
6548 & [N~II]               &  0.20            & ----        &   ---- & ----\\
6563 & \Halpha          & 2.90             & 2.66      & 1.00 & 1.00\\
6583 & [N~II]               & 0.66              & 0.68     & 0.23  & 0.23\\
6716 & [S~II]               & ---- & ---- & 0.13 & 0.16\\
6716+6731 & [S~II]    & 1.00              & 0.79& ---- & ----\\
\hline
\end{tabular}
\end{minipage}
\end{table*}

\subsection{Earlier Observations of the Orion-Eridanus Bubble}

The \citet{mad06} study also included high signal to noise ratio fixed-pointing samples at four additional positions to the east of the center of OB stars in the Belt and Sword regions of the nebula, extending
out to a distance of 24\fdg0. They also obtained seven closely spaced samples crossing an ionization front at the southwest boundary of the \bub\ and lower quality data in regions  further from the Galactic Plane and outside but near the \bub, thus sampling the more general WIM.  When they compared their results with a similar set of measurements of large and low density H~II regions they noted that the [S~II]/\Halpha\ and [N~II]/\Halpha\ flux ratios for the 
\bub\ varied together in the same way as in the WIM and they concluded that the variations in flux ratios were due to variations in \Te , with the WIM being significantly warmer than the \bub\ material. They established that the pattern of the line ratios was quite different for the H~II regions they had sampled and the \bub\ and WIM. The difference was in the sense that the [S~II]  to \Halpha\ ratio was much weaker in the H~II regions at the same value of the [N~II]  to \Halpha\ ratios. Moreover, they noted that the [S~II] 6716 \AA\ over [N~II] 6583 \AA\ flux ratio was both larger than in their H~II regions and increased only slightly with surface brightness in \Halpha, with an average value near 0.8. The value of this ratio was about  0.4 in their low surface brightness H~II regions.

\section{Modeling}

It is important to determine the physical conditions in the Barnard's Loop and the \bub\ in order to understand both the nature of these objects and their origin.  As we will see, there is a fundamental problem when trying to explain the line ratios (which reflect the conditions of ionization and excitation) with the mechanism of direct stellar photoionization by the hottest star in the region, but, we have been able to develop a model consistent with all the available observational material.

There are two methods of testing the source of illumination of Barnard's Loop. The first test is to determine if the total stellar luminosity in the LyC that is derived from the surface brightness is consistent with the available flux from the enclosed stars and is considered in \S\ 3.2. The second test is to determine if the observed emission-ine  ratios can be explained by photoionization caused by the enclosed stars and is considered in \S\ 3.4 for the well studied object M~43 as a test object and then in \S\ 3.5 we apply this method to the Barnard's Loop.

\subsection{Basic Constraints Identified from the Observations}

The observations of the emission distribution, the surface brightness, and the known distance allow us to  determine some basic properties of Barnard's Loop.  Fortunately the interstellar extinction in the region is known to be negligible \citep{pei75,mad06}, a conclusion consistent with the fact that our new \Halpha/\Hbeta\  flux ratio of 2.90 is that predicted for an unreddened gas of about 9000 K.  

In their fundamental study of the Orion Nebula, \citet{b91} demonstrated that the surface brightness in the \Hbeta\ recombination line along a line of sight from the observer to the photoionizing star would be proportional to the flux of ionizing photons $\phi$(H) in units of photons cm$^{-2}$ s$^{-1}$ reaching a thin slab of gas. This approximation is probably valid for Barnard's Loop because the low inner-region densities would not be expected to absorb LyC photons. If one views the emitting layer face-on, then the relation would be $\phi$(H)=4$\pi$($\rm \alpha _{B}$/$\rm \alpha ^{eff}_{H\beta}$)S(\Hbeta) where the surface brightness S(\Hbeta) is expressed in \sbu\ and $\rm \alpha _{B}$ and $\alpha ^{eff}_{H\beta}$ are respectively the total recombination coefficient for hydrogen and the effective recombination coefficient for the \Hbeta\ line. The ratio of the recombination coefficients is not very sensitive to the assumed \Te .  

In the case of Barnard's Loop we are observing a relatively thin shell seen nearly edge-on, so that the surface brightness will be enhanced by a factor of g above that for a slab viewed face on. We call the quantity  the limb-brightening correction. In this thin shell model the surface brightness distribution would start at the maximum radius R$\rm _{max}$, rapidly rise to a peak value at angle corresponding to the inner radius R$\rm _{peak}$, then slowly decrease to the value that would apply along the line of sight through the ionizing star. The enhancement of the surface brightness will be g=2( R$\rm _{max}^{2} $-R$\rm _{peak}^{2}$)$^{1/2}$/(R$\rm _{max}$ - R$\rm _{peak}$). Using the dimensions for the line from \ori\ through our CTIO-2008 observations (R$\rm _{max}$=7\fdg07 and R$\rm _{peak}$=6\fdg01), the geometric correction factor is g=7.03, that is, the observed peak surface brightness will be enhanced by a factor of 7.03.  

One can derive the flux of ionizing photons ($\phi$(H)) from S(\Hbeta) after the former has been corrected for limb brightening.  Using our observed surface brightness ($5.3\times 10^{6}$ \sbu), the geometric correction factor of 7.03 yields a flux of ionizing photons $\phi$(H)=8.1 x 10$^{7}$ photons cm$^{-2}$ s$^{-1}$. It is hard to ascribe a probable error to this number. The observational uncertainty is less than 10\%. The uncertainties in the derived results that are due to applying a simple thin-shell model to Barnard's Loop, which has some internal structure, are probably significantly larger.  As we will see in the next section, this flux and its wavelength distribution determine the expected photoionization structure and observed line ratios within Barnard's Loop. Having calculated the ionizing flux in a given sample, one can calculate the total luminosity in LyC photons  (Q(H) with the units photons s$^{-1}$) for the ionizing star(s) by the relation Q(H)=4$\pi$ R$^{2}$$\phi$(H). Adopting the average of the maximum and peak surface brightness radii of $1.56\times 10^{20}$ cm yields Q(H)=$2.5\times 10^{49}$ photons s$^{-1}$. 

This Q(H)  value is larger than the value for the Orion Nebula of $7.8\times 10^{48}$ photons s$^{-1}$ found by \citet{pei75} from early radio continuum observations and the value for the Orion Nebula of $1.1\times 10^{49}$ found by \citet{vdw89} from their VLA study. The Orion Nebula Q(H) values are in approximate agreement with those expected from dominant photoionization of that object being by \ori\ and \oriA, where the expected Q(H) for the stars are $6\times 10^{48}$ photons s$^{-1}$ and $1.5\times 10^{48}$ photon \pers\ respectively, using the calibration of \citet{srh06}. The fact that the predicted stellar value of Q(H) is smaller than that derived for the nebula probably indicates the uncertainty in the calibration or that the true spectral type is slightly earlier. 

One can also estimate the electron density from a knowledge of the surface brightness, the emissivity in the \Hbeta\ line and the geometry. For the constant density shell model the electron density ($n\rm _{e}$) will be $n\rm _{e}^{2}$=4$\pi$ S(\Hbeta)/[$\rm \alpha ^{eff}_{H\beta}$ g(R$\rm _{max}$-R$\rm _{peak}$)], assuming that all the free electrons arise from the photoionization of hydrogen, a very good approximation in this low-ionizationization region where helium is neutral (as indicated by the very weak or absent HeI 5876 \AA\ line). Using the previous values and adopting $\rm \alpha ^{eff}_{H\beta}$=$3.63\times 10^{-14}$ cm$^{3}$ s$^{-1}$ from \citet{agn3} yields $n\rm _{e}$=3.2 \cmq\ for the region of Barnard's Loop that we have observed and 0.7 \cmq\ for the \citet{mad06} western arc. These numbers are similar to the density $n\rm _{e}$=2.0 \cmq\  derived by \citet{carl00} from combined optical and radio data over large samples in Barnard's Loop.

 \subsection{Comparison of the Derived LyC Luminosity with that of the Enclosed Stars}

A natural first step in explaining the observed emission line properties is to investigate whether the line ratios can be explained by photoionization from a single dominant star, an approach adopted qualitatively by \citet{pei75}. We will follow their discussion except that we shall use the spectral types summarized in \citet{gou} and use the Q(H) values from the recent study by \citet{srh06}. There are many candidate ionizing stars within Barnard's Loop and its extension into the \bub,  the brightest  six in the LyC are the brightest Trapezium O6~V star \ori\  with an expected LyC luminosity of 6 x 10$^{48}$ photons \pers, the cooler (spectral type O9.5 II) $\delta$\ Ori with an expected LyC luminosity of $5.6\times 10^{48}$ photon \pers\ , $\zeta$\ Ori (O9.5 Ib) at $5.6\times 10^{48}$ photon \pers\ ,  and  $\iota$\  Ori (spectral type O9~III) at $6.6\times 10^{48}$ photon \pers , plus  \oriA\ and $\sigma$ Ori (both 09.5~V and $1.5\times 10^{48}$ photon \pers). It is unclear if $\epsilon$\ Ori belongs in this list. With a spectral type of B0~Ia it is cooler than the stars studied by \citet{srh06}. \citet{vacca} did include one star (HD~37128) of this spectral type in their study and found it to be be 6000 K cooler than an O9.5~Ia spectral type star (HD~30614) and would therefore have a much lower LyC luminosity. Since $\zeta$\ Ori has a spectral type of O9.5~Ib, we conclude that $\epsilon$\ Ori's LyC luminosity is much lower than $5.6\times 10^{48}$ photon \pers\ and we will not include it in our tally. The location of these stars are shown in Figure 1. Although there are many additional stars of later spectral type and less luminous stars are distributed throughout the inner Orion constellation, their contribution to photoionization of Barnard's Loop must be minimal. For a summary of the properties of these stars see Table 1.1.IV  of \citet{gou}.  

The  \Hbeta\ surface brightness  yields an ionizing flux of $\phi$(H)=$8.1\times 10^{7}$ photons cm$^{-2}$ s$^{-1}$\ that then leads to a derived ionizing luminosity of Q(H)=$2.5\times 10^{49}$ photons s$^{-1}$. Photoionization of the large scale Barnard's Loop may arise from multiple stars. It is unlikely that two candidate stars that are located within the Orion Nebula (\ori\ and \oriA) are important contributors to ionizing Barnard's Loop because it is well known \citep{od01} that the nebula is optically thick to LyC radiation in all directions except possibly to the southwest, where the foreground Veil thins and X-ray emission from million degree gas is found \citep{gud08}.The four remaining candidates for causing photoionization would have a total luminosity  of Q(H)=$1.9\times 10^{49}$ photons \pers, in reasonable agreement with the LyC luminosity derived from the \Hbeta\ surface brightness.  The effective temperatures of these stars range from 30000 K to 32000 K. 
 
\subsection{Method of Calculating Photoionization Models}

We have determined the properties of the radiation field illuminating Barnard's Loop by comparing the predictions for emission line ratios from stellar atmosphere models of various effective temperatures (\Tstar ). We use the development version of the spectral simulation code Cloudy, last reviewed by \citet{fer98} and noted in appendix A. We present here several large grids of model calculations that, in some cases, varied both \Tstar\ and the flux of hydrogen-ionizing photons, and in others, only \Tstar . Later we also calculate and present models including differences in 
the abundance ratio (Z/H) of heavy elements to hydrogen. In many ways our calculations are similar to those of \citet{sem00} but we use a version of Cloudy with up-to-date atomic coefficients, more recent stellar atmosphere models, and more recent calibrations of the spectral type-\Tstar\ relation.  

The other key parameters in the calculations are the abundances of the heavy elements that dominate the cooling of the ionized gas. We adopted  N/H=$6.5\times 10^{-5}$, O/H=$4.3\times 10^{-4}$, and S/H=$1.4\times 10^{-5}$ for these as a point of reference, using these for the M~43 calculations, where these abundances are known to apply, and scaled the abundances for the Barnard's Loop and WIM calculations, where there are no direct determinations of the abundances. The O/H and N/H ratios are  averages of the similar interstellar medium \citep{jen09} and M~42 \citep{sim04} results and the S/H ratio of $1.4\times 10^{-5}$ was taken from the study of \cite{daf09}, who measured two stages of ionized sulphur in ten B stars in the Orion Association.  We also adopted a ratio of extinction to reddening of R=3.1, the standard dust to gas ratio, and included PAH's, and the TLUSTY stellar atmospheres of \citet{lanz}. Density is a less important factor, but we have used the appropriate values for the two specific objects being modeled.

We must note that the S/H ratio derived from H~II regions can be significantly smaller than the solar abundance of S/H =$1.4\times 10^{-5}$ \citep{asp06}. \citet{sim04} give S/H=$7.0\times 10^{-6}$ from IR lines in the Orion Nebula, a value consistent with previous optical studies.  However \citet{daf09} find that Orion B stars have S/H close to the solar value, and the \citet{est} study of optical emission lines find relatively high values for the Orion Nebula ranging from $1.1-1.7\times 10^{-5}$. Jenkins (2009) finds S/H depleted below the solar value in the ISM. These S/H values range over a factor of two, and Jenkin's (2009) discovery of a depletion pattern in the ISM means that S/H may depend on the location in the Galaxy.  This affects Figure 5.  The regions where [S~ II] lines form are mainly cooled by lines of [S~II], [N~II], and [O~II].  We have adopted S/H = $1.4\times 10^{-5}$. Had we used a lower value the [S~II] lines would be weaker, but by less than the change in the S/H ratio because of the thermostat effect in a photoionized gas \citep{agn3}.  If the [S~II] cooling is diminished by lowering S/H, the gas will grow hotter, making lines of [O~II] and [N~II] stronger.  The effect would be to shift the curves in Fig 5 down and to the right.  A factor of two decrease in S/H, which would be consistent with the older emission line studies, would shift the log $\phi$(H) = 6.7 curve to roughly the position of the  \citep{mad06} data on this plot. A similar large range in N/H abundances was found by \citet{sim95} in their study of galactic H~II regions, something that becomes important in our discussion of the indirect method of determining \Te\  in WIM components. These known abundance variations play an important role in our interpretation of line ratios of both the Barnard's Loop and the WIM, as discussed in \S\ 3.5 and \S\ 3.6. Our abundance ratios most closely resemble the B star abundances adopted by \citet{sem00} except that we assume a higher overall abundance.  

The interstellar medium on the galaxy-wide scale is likely to be inhomogeneous and porous Ð that is, certain regions may be relatively clear of material, and others affected by nearby dense clouds \citet{spi78}.  Quite sophisticated models have been developed to take this geometry into account (see, for example,  \citet{wood05,haf09}).  Although Cloudy is capable of simulating complex geometries (the code ÒCloudy 3DÓ, see \citet{mos08} and http://sites.google.com/site/cloudy3d/), here we use it to simulate a shell of gas that is symmetric about the central star or star cluster. The temperature, ionization, and emission in many thousands of lines are determined as a function of distance away from the central ionizing stars.  Line and continuum optical depths are taken into account so that changes in the ionizing radiation field (Chapter 3, \citep{agn3}) are reproduced and the geometry is an inhomogeneous mix of temperature, ionization, and emission.  In this mode a shell of gas, which appears to describe both the Barnard Loop and M~43, is closely mimicked.  The quantities we report correspond to the observed spectrum along a tangent occurring at various impact parameters through the shell away from the star cluster.  In this way we closely mimic the observations, which place a spectrometer slit at various positions.

There are at present no reliable dielectronic recombination rate coefficients for recombination from S$^{+2}$ to S$^{+1}$.  We reply on empirical estimates of the rate coefficient based on photoionization modeling of astronomical observations, as was done by \citet{ali91}.  Based on such models we would judge our current estimates to be uncertain by about 30\%.  This introduces roughly a 20\%\ uncertainty in the balance between [S~ II] and [S~III].  This produces an uncertainty that enters as an unknown scale factor that affects all models. This represents a basic uncertainty that affects all observations by this scale factor.  It may shift the model predictions by an unknown systematic amount, but will not produce object to object fluctuations in the spectrum.

\subsection{Derivation of the Ionizing Star Properties from Emission Line Ratios in M~43}

To demonstrate the validity of the method employed in this paper to Barnard's Loop, we first made calculations that simulate the H~II region M~43 (NGC~1982), which is dominated by a single cool star. We then used the results of that study to guide the  calculations of the conditions for Barnard's Loop. 

M~43 lies to the immediate northeast of the much brighter Orion Nebula (M~42, NGC~1976) and was included in the recent comprehensive spectroscopic mapping of \citet{od10}. The central star is NU Ori, which is spectral type B0.5~V \citep{sh71,pen75,sd08}. The total Lyman continuum luminosity of the nebula is about $4.7\pm2.0\times 10^{47}$ photons s$^{-1}$ as determined from multiple radio wavelength studies summarized by \citet{gou}.
 The apparent radius is 128\arcsec\ and the distance is 440 pc,  giving a true radius is 0.27 pc. If the radius is adopted as the distance between the central star and a blister-type ionization front as applies to M~42, then the corresponding $\phi$(H) value is $5.3\times 10^{10}$ photons \cmsq\ s$^{-1}$. We have adopted an electron density in M~43 of 520 \cmq\ \citep{od10}.  We do not use the ionizing luminosity expected from the spectral type of NU Ori because it is at least a triple star system \citep{mor91,prei99} and is a known windy star with a constant strong X-ray flux \citep{stel05}. Therefore, we take advantage of knowing the total luminosity of the nebula and leave the temperature of the ionizing star as a variable.

In this case we assume that the abundance of elements is that of Section 3.3 but  that we don't know the exact geometry (the apparent radius need not be the same as the separation between the central star and the ionization front), so that we have varied the adopted $\Phi$  values for a series of models for stars of varying \Tstar\  (27542 K through 40040 K in steps of 0.0325 dex), all with stellar gravities corresponding to main sequence stars. The results are shown in Figure 3. To these color-color diagrams we have added the observed line ratios for the six samples in M~43 from the \citet{od10} study that gave I([N~II] 6583 \AA)/I(\Halpha)=0.41$\pm$0.02, I([S~II] 6716 \AA+6731 \AA)/I(\Halpha)=0.18$\pm$0.03, and I([O~III] 5007 \AA)/I(\Hbeta)=0.26$\pm$0.11. 

Examination of the low-ionization color-color diagram (Figure 3-left) shows that the best agreement of the emission line ratios with the predictions of the model stellar atmospheres is for a star with \Tstar\ between 29700 K and 34500 K and  log $\Phi$$\simeq$11.0. These temperatures bracket the temperature of $\sim$30000 K expected for a spectral type B0.5V  \citep{srh06} and the best log $\Phi$ value indicates that the distance between the ionizing star and the ionization front is 0.7 times the apparent radius of the nebula.
A B0.5~V star would be expected to have a LyC luminosity \citep{agn3} of $8\times 10^{47}$ photons s$^{-1}$, within a factor of two of the luminosity derived from the radio continuum. 

Examination of the high-ionization color-color diagram (Figure 3-right) shows that the I([O~III] 5007 \AA)/I(\Hbeta) observations are best fit by a much higher \Tstar . The cause
of this is probably not in the models, rather that the line ratio is contaminated by light originating in the nearby Huygens Region, where this ratio is about 3 \citep{od10}, and it would only take about 20 \%\ contribution to I([O~III] 5007 \AA) to account for the discrepancy. \citet{od10} argue that
essentially all of the ionization of M~43 arises from NU Ori, but their analysis does not preclude this level of contamination by scattered light. At similar distances within M~42 the scattered light from the Huygens Region is about this level.

We show in Figure 4 the results of calculations for predictions of  \Te\ for the most likely value of log $\Phi$ determined from Figure 3-left. We have superimposed the average \Te\ derived in \citet{od10} from the nebular to auroral transitions of [N~II] (7950$\pm$720 K) and [S~II] (7260$\pm$720 K). The uncertainties in the temperatures (because of the weakness of the auroral transitions) is too large to narrowly define the stellar models, but we see that again the \Te\ taken together agree with \Tstar\  derived from the spectral type.

\subsection{Derivation of a Model for Barnard's Loop from the Ratios of the Strongest Emission Lines}

We then performed a similar set of calculations for Barnard's Loop, using the parameters relevant for that object. Fortunately, the constraining 
factors are better known than for M~43.
In this case we know that all of the candidate ionizing stars ($\delta$~Ori-09.5~II), $\zeta$~Ori-09.5~Ib, $\iota$~Ori-O9~III, $\sigma$\ Ori~O9.5~V) fall in a narrow range of \Tstar\ near 31000 K. The side-on view of Barnard's Loop means that we know the separation of the ionizing stars and the ionization front  reasonably well and in \S\ 3.1 derive an ionizing flux of $\Phi$=$8.1\times 10^{7}$ photons \pers. We also know that the density is about 3.2 \cmq\ for our sample. Our calculations were made using \Tstar =31000 K and a stellar gravity appropriate for a giant star. We adopted the previously assumed properties of interstellar grains and PAH and initially used the same abundaces as in our calculations of M~43 models.

In these new calculations there was a significant disagreement between the predicted model and the observations in both the low-ionization color-color (Figure 5) and the high-ionization color-color (Figure 6) diagrams. This disagreement was present whether one uses all of the available observations or only the ratios from this study. We tried to fit models with different values of $\Phi$ and \Tstar\ as low as 27500 K, but none of these agreed with the observations. The locus of points for various values of log~$\Phi$ for the closest low \Tstar =27500 K models is also shown in Figure 5 . The small discrepancy at 27500 K becomes larger at higher values of \Tstar\ since plots for \Tstar\ progress to the right in the diagram, with the \Tstar\ matching the likely ionizing stars shown as an open square around the a filled circle (log~$\Phi$=7.9).
  
 We then explored the results of varying the only parameter in our set of assumptions that is not directly determined for the Barnard's Loop, that is, the relative abundances of the heavy elements (Z/H).  We scaled all of the heavy elements together,  thereby ignoring subtleties in things like the N/O ratio changes that can be expected from stellar evolution models.  In this case we see that one obtains excellent agreement between the observations and 
the models with a log~(Z/H) abundance enhancement of between 0.1 and 0.2, i.e. a heavy element abundance enhancement of about a factor of 1.4.

The more diagnostically useful figure is the low-ionization color-color ratio because of its insensitivity to contamination by scattered light and Figure 5 will be used for discussion. It is important to understand that the locus of points tracked by the models of various abundances is a result of both the number of elements of that species available for emitting the emission line, but also \Te\ (the emissivity of the collisionally excited [S~II]  and [N~II] lines increasing with \Te\ and the emissivity of the recombination \Halpha\ decreasing with \Te ). A track from lowest Z/H to highest Z/H models is a monotonic progression from high to low \Te .

Comparison of our models with varying abundances with the three sets of observations of Barnard's Loop (this study, \citep{pei75}, \citep{mad06}) still indicates that the best agreement is when there is an enrichment of heavy elements by between 0.1 and 0.2 dex.

We also considered the high-ionization color-color diagram because the [O~III] 5007 \AA\ line was detected by \citet{mad06} in several of their 
samples of Barnard's Loop.  In each case there was significant disagreement with the models best fitting the low-ionization color-color diagram in the sense that there was excess [O~III] emission. In the same way as M~43, this excess is almost certainly due to scattered light from the bright central H~II regions in Orion where the I([O~III] 5007 \AA)/I(\Hbeta) ratio is about three and a contamination of only a few percent is enough to account for the disagreement. 

As a guideline for future observations of Barnard's Loop, we present in appendix B the predicted relative intensities of the strongest emission lines over a wide range of wavelengths. We adopted \Tstar\ = 31000 K, log~$\Phi$ ~=~7.91, and an abundance enrichment of 0.15 dex as the best fit to the Barnard's Loop.

 \subsection{Comparison of Barnard's Loop Properties with those of the \bub\ and the WIM}
 
 \citet{mad06} measured a large number of samples within and near the \bub\, in addition to two samples (W1 and W2) that clearly lie within Barnard's Loop. They point out the continuity of the low-ionization color-color diagram results for all of these, in spite of a factor of 40 range of \Halpha\ surface brightness. Their results for the low surface brightness H~II regions fall below the pattern established for the other samples.  In Figure 7 we show the data from Madsen et~al.'s (2006) Figure 7 and have added our new Barnard's Loop observations and those of \citet{pei75}.  The similarity of the Barnard's Loop ratios (especially our own) and those in the \bub\  and the WIM samples is striking and probably indicates a common set of physical conditions and processes.
 
 To these observations we have added the predictions for our models used to explain the Barnard's Loop observations shown in Figure 5. The log~$\Phi$=7.9 calculations have been designated  by the more general term, the ionization parameter U (log~U=-3.07).  U is the ratio of the number density of ionizing photons and hydrogen atoms and is given by U=$\phi$/(c~n$_{H}$), where c is the velocity of light and $n_{H}$ is the total hydrogen density. U is a broadly useful parameter because it directly indicates the conditions for photoionization, with various values of flux and gas density producing the same conditions for photoionization.  U becomes a less useful measure in considering bright and dense photoionized gases because it does not give an indication of collisional processes. The same coding of log~ (Z/H) has been used for both Figure 5 and Figure 7 .
 In addition to the set of models displayed in Figure 5, we also show a series of calculations using our  hottest stars at 40000 K (corresponding to a spectral type of about 07.5) at log~U=-3.07 and stars of 34500 K and 40000 K for log~U=-3.67.
 
 Examination of Figure 7 indicates that variable abundance models for log~U=-3.67 and \Tstar =3100 K, log~U=-3.67 and \Tstar =40000 K, and log~U=-3.07 and \Tstar =31000 K
 fully enclose the space occupied by the Barnard's Loop, the \bub, and the WIM sample observations, with only the low surface brightness H~II region observations falling below. The region occupied by the low surface brightness H~II regions demands higher values of U than the Barnard's Loop although comparable values of Z/H and \Tstar. Within the envelope of these bounding models one cannot tell from the low-ionization color-color diagram  of the WIM samples what 
 causes an individual point to have its specific location, since the location is defined by U, \Tstar , and Z/H. In the case of the Barnard's Loop samples, where the Z/H and \Tstar\ values must be nearly constant, we'd expect variations in U to distribute the observed points along a line of constant color (a nearly vertical line) and that is the case.  The wider distribution of the other \bub\ samples  (the darker Madsen et~al. (2006) points in Figure 7) would then indicate variations with position of abundance and/or \Tstar\  of the dominant ionizing star, in addition to variations in U.  Variations in \Tstar\ are certainly possible in such a large-scale sampling since there is evidence \citep{od01} that the optically thick foreground Veil of the Orion Nebula is probably optically thin to the southwest  and this would allow radiation from the hottest star in the region to illuminate \bub\ components in that direction. In the case of the WIM samples it is expected that there could be a significant range in photoionizing star temperatures, U, and possibly Z/H.
 
 If there is only a single value of the (Z/H) and it is that adopted for our M~43 calculations then most of the \bub\ and WIM ratios can be 
 explained by log ~U values between -3.07 and -3.67, with \Tstar\ values of up to slightly more than 35000 K. However, the lower left population
 of the \bub\ samples and the Barnard's Loop samples would require unrealistically low \Tstar\ values, indicating that there must be regions of higher than average Z/H.
 
 We can constrain the likely \Te\ of the Barnard Loop samples since they are all illuminated by the same radiation field.  Table 2 gives the \Te\ in the [N~II] emitting zone for all our models. The two most closely matching the low-ionization color-color diagram are those with log~U=-3.67 and log~U=-3.07 with \Tstar =31000 K, both with an abundance difference of 0.1 dex, and these have expected \Te\ of  5970 K and 5940 K respectively. We will adopt a value of 5960$\pm$50 K for comparison with direct determinations. There is a great uncertainty about the expected \Te\ of the other parts of the diagram. For example, the \Tstar=31000 K and log~U=-3.07 model with average Z/H predicts line ratios about nearly the same as the \Tstar =40000 K and log~U=-3.67 model with an abundance enhancement of about 0.5 dex, would have the very different temperatures of 6530 K and 4680 K.
 
The important conclusion of this section is that theoretically one can explain the low-ionization color-color diagram for the non-Barnard's Loop samples by a range of values of U, \Tstar , and Z/H. A simple ratio of nebular line intensities for two different ions cannot produce an unambiguous estimate of \Te .

 \begin{table*}
 \centering
 \begin{minipage}{126mm}
 \caption{Predicted Electron Temperatures in the [N~II] Emitting Zone.}
 \label{tabletwo}
 \begin{tabular}{@{}ccccc@{}}
 \hline
----- & Log~U=-3.67     & log~U=-3.67       & log~U=-3.07          & log~U=-3.07\\
 $\Delta$~log~(Z/H) & \Tstar=31000 K & \Tstar=40000 K & \Tstar=31000 K & \Tstar=40000 K\\
 \hline
 -0.5   &  8950 &10280 & 9270 & 10940\\
-0.4   &  8490 & 9890 & 8730 & 10420\\
-0.3   &  8020 & 9440 & 8200 & 9860\\
-0.2   &  7530 & 8950 & 7660 & 9270\\
-0.1   &  7010 & 8410 & 7100 & 8650\\
0.0    &  6500 & 7850 & 6530 & 8020\\
0.1    &  5970 & 7260 & 5940 & 7360\\
0.2    &  5420 & 6670 & 5330 & 6680\\
0.3    &  4850 & 6040 & 4700 & 5970\\
0.4    &  4210 & 5380 & 4000 & 5240\\
0.5    &  3480 & 4680 & 3280 & 4440\\
\hline
\end{tabular}
\end{minipage}
\end{table*}
 
 \subsection{Direct Determinations of the Electron Temperatures in H~II Regions, Barnard's Loop,  and the WIM}
 
 The reason for the designation WIM (warm ionized medium) is the fact that \Te\ there is higher than the cold gas in the ISM. The actual value for the WIM's temperature is much more uncertain than sometimes stated in the literature because there are few observations (summarized in this section) that allow a direct determination and the indirect methods (based on only the I([N~II] 6583 \AA)/I(\Halpha) ratio) commonly employed are uncertain as they assume a fixed nitrogen ionization ratio and abundance. In this section we summarize the results for \Te\ derived by direct means.
 
There are three direct methods of determining \Te. The first is from the measurement of forbidden line intensity ratios within a single ion. The second is from the width of emission lines from ions of very different mass. The third is from the ratio of continuum to recombination line emission.

Observations of the low-ionizationization WIM cannot use the most widely used \Te\ indicator, the [O~III] auroral/nebular line ratios. However, \citet{rey01} were able to measure the auroral 5755 \AA\ and nebular 6583 \AA\ lines of [N~II] in a number of low surface brightness H~II regions and several samples of WIM clouds along a single line of sight. \citet{mad06} were able to measure these same line ratios in several additional WIM clouds and WIM clouds lying along the same line of sight as low surface brightness H~II regions. They conclude that the WIM component has a line ratio corresponding to \Te\ about 2000 K higher than their sample of H~II regions and that \Te\ is higher for WIM clouds of lower surface brightness.

 We show the \citet{rey01} and \citet{mad06} data in Figure 8 and have added line ratios for many additional brighter H~II regions \citet{gar04,gar05,gar06a,gar06b,od10}.   The \citet{rey01} WIM sample had four velocity components and we see that when considered together (the ``total'' sample) the line ratios are similar or lower than most H~II regions, indicating similar \Te .  Only the WIM component along the line of sight to Sivan 2 has a significantly higher inferred \Te\ and it is comparable to the hottest classical H~II region (NGC~3603). In contrast, the low surface brightness H~II regions are systematically lower in line [N~II] line ratio  than the bright H~II regions,  hence have lower \Te . The H~II regions in the \citet{rey01} sample have \Halpha\ surface brightnesses of 68--339 R. They are much lower surface brightness than classical bright H~II regions. For reference, the Orion Nebula has an extinction corrected maximum surface brightness in \Halpha\ of $1.1\times 10^{6}$ R \citep{od10}.
 
The conclusion that one can draw from the data presented in the \citet{rey01} and \citet{mad06} studies and comparison with the results from classical bright H~II regions is that the only direct determinations of \Te\ of WIM components indicate temperatures comparable to classical bright H~II regions. It is only the low surface brightness H~II regions in the \citet{rey01} and \citet{mad06} sample that are of unusually low \Te, in spite of the emphatic statement in \citet{haf09} that the WIM temperatures are elevated to the bright classical H~II regions. That statement is only true for the very low surface brightness components of the WIM, as we discuss in \S\  4.3. This can be due to them being photoionized by cooler stars. An additional factor may be that collisional de-excitation is not important at the lower densities in the WIM clouds so that cooling radiative transitions are relatively more important. There is an observational selection effect in the WHAM studies in that the low surface brightness H~II regions are much larger than the typical H~II regions since they are usually larger than the 1\degr\ diameter of the WHAM field of view. 
 
Line widths will characteristically have several components, the thermal width (which will scale as the square root of the ratio of temperature and ion mass) and random large scale mass motions along the line of sight.  Since the atomic mass of S is 32 times that of H, one can hope to determine \Te\ after making reasonable assumptions about the common properties of the mass motion.  In an early WHAM study, \citet{rey85} compared the line widths of \Halpha\ and [S~II] lines, but the spectral resolution and intrinsic widths of the lines did not allow an unambiguous determination of \Te.  Line widths have been measured in the diffuse H~II region surrounding $\zeta$~Oph, where the fainter components are comparable to the WIM in surface brightness.  Observations of [N~II] 6583 \AA, [S~II] 6716 \AA, and \Halpha\  were interpreted by \citet{baker04} in a paper that only appears as an abstract. However, a summary of their results is shown as Figure 4 of  \citet{haf09}.  There one sees that most of the observed points lie between 6000 K and 9000 K and that there is a systematic and position dependent change in the non-thermal component.  Again these temperatures are comparable to the classical bright H~II regions.

Comparison of optical emission lines and radio continuum can also provide a good source of \Te.
In the case of Barnard's Loop, \citet{carl00} use their own radio observations in two large samples at four wavelengths and WHAM optical surface brightnesses to determine that \Te\ is about 6100 K, in excellent agreement with the value of about 5960$\pm$50 K inferred from the low-ionization line ratios discussed in \S\ 3.6. 
An unusual result  using this approach occurs for the WIM . \citet{dobler} compared the continuum measured with the WMAP satellite with the \Halpha\ surface brightness. They determined that the observed ratio required a gas temperature of 3000 K.  It is difficult to understand such a low temperature arising from photoionization processes, which has lead to the creation of a three component model composed of photoionized gas, gas that is recombining and cooling, and cool neutral hydrogen \citep{dong}. Following \citet{wood99}, \citet{dong} assumed in their model that there was a significant scattered light component, and found that a 15\%\ scattered light contribution to \Halpha\ was necessary to produce a satisfactory model. It is hard to assess this multi-free-parameter model since it assumes that the warm gas component temperature is that indicated by the I([N~II] 6583 \AA)/I(\Halpha) ratio method. There are also arguments based on direct observations that high latitude clouds are strongly affected by scattered \Halpha\ radiation \citep{witt}. If contamination by scattered \Halpha\ was stronger than derived in the detailed models of \citet{dong}, then the temperature of the WIM gas contributing to the radio continuum would be higher than 3000 K.

\section{Discussion}

In this section we consider the results derived from the earlier sections. We consider the effects of variations of Z/H on the low-ionization color-color diagram, the general utility of using the low-ionization color-color diagram for determining \Te, the systematic changes of \Te\ determined from the low-ionization color-color diagram, and the issues involved with using large spatial scale line ratio variations from observations of other galaxies.

\subsection{Effects of variations of Z/H on the low-ionization color-color diagram}

Our most important conclusion is that one cannot explain the low-ionization color-color diagram for the Barnard's Loop samples by the stars
capable of causing this large-scale photoionization if the Z/H ratio is our reference value used in the M~43 calculations. We invoke a 
heavy element enhancement of about a factor of 1.4 to reconcile our models with the observations.  A similar enhancement is also necessary
to explain the high surface brightness population of the \bub\ samples. If such variations are necessary to explain the best studied samples, it is likely that the abundance variations also occur in the clouds producing the components of the WIM population.  

There certainly is evidence for local variations in Z/H within the ionized components of the interstellar medium. Two studies of multiple H~II regions and compact H~II regions at various galactocentric distances \citep{sim95,aff97} found a general decrease in Z/H with increasing distance from the center of the Galaxy.  Their analysis relies on ratios of infrared emission lines and will not be affected by uncertainties in the gas temperature or density. More 
important to our problem, they found in the local part of the Galaxy, variations of 0.3 dex both above and below the average and well beyond
their estimated probable errors. These variations probably arise because stars with a range of masses and lifetimes produce different elements that enhance their locale.

As noted in \S\ 3.5, we have assumed in our models that the relative abundance of various elements vary together. For a small-scale object, where the products of an individual evolved star can be important, this assumption would be questionable. However, for the large objects that we consider here (Barnard's Loop, the \bub\, and the WIM clouds) the abundances must be determined by the products of multiple stars and the adoption of a constant ratio of individual elements is justified.

\subsection{Application of the low-ionization color-color diagram for determining \Te}

During the last decade there have been multiple papers on the WIM that argue that a low-ionization color-color diagram similar to Figure 7 can be explained  primarily by variations in \Te\ as noted in the recent review by \citet{haf09}. The method goes back to the WIM study by \citet{haf99} and has frequently been employed \citep{mad06}.  Within the assumption that the ionization ratio of nitrogen remains constant and a known N/H abundance applies, it is a simple matter to draw vertical lines representing constant values of \Te\ in Figure 7. The nitrogen ionization ratio is defined as (H/H$^{+}$)(N$^{+}$/N), where H and N represent the total number density of hydrogen and nitrogen atoms and the superscripted values their ion number density. This reflects the fact that the emissivity of a recombination line and forbidden line have reverse-sense dependencies on \Te . However, if the nitrogen ionization ratio is lower than assumed the relative emissivity per unit volume decreases and a vertical line indicating a fixed \Te\ would move to the left. If the N/H ratio is lower than assumed, the displacement would also be to the left. Of course both assumptions could be incorrect  in different senses and one error can correct for the other, but without a good knowledge of both the nitrogen ionization ratio and the relative abundance of nitrogen and hydrogen, the method is suspect.

In an attempt to identify the range of probable values of the nitrogen ionization ratio, we have extracted this information from our calculated models. 
For the models that most closely match the distribution of the Barnard's Loop, \bub, and WIM observations the ionization ratio varies little. The nitrogen ionization ratio for log~U=-3.67 and \Tstar=31000 varies only from 1.007 to 1.014 over the range of Z/H from -0.5 to 0.5 dex. The ionization ratio for log~U=-3.07 and \Tstar=40000 varies only from 1.012 to 1.035 over the range of Z/H from -0.5 to 0.5 dex.  This confirms that variations in the ionization ratio do not play an important role, as previously assumed and calculated. This is in excellent agreement with the predictions of Sembach et~al. (2000). Unfortunately, in the study of \cite{mad06}, which drew on the Sembach et~al. (2000) models a value of the nitrogen ionization ratio of 0.8 was adopted, whereas this is actually the value for N$^{+}$/N. This error was not corrected when the results were repeated in a review article \citep{haf09}.  This makes the electron temperatures they present to be too large, for their assumed abundance. In the original study of \citet{haf99} a nitrogen ionization ratio of 1.0 was adopted, which means that those temperatures should be correct, if the nitrogen abundance they adopted of N/H=7.5 x 10$^{-5}$ is both correct and uniform.

The more important limitation of the I([N~II])/I(\Halpha) ratio method is the other scaling factor, the N/H ratio. We argue in \S\ 3.6 that in the case of 
Barnard's Loop that there is a Z/H abundance enhancement of about 0.15 dex that (alone) would shift a line of fixed \Te\ to the right
by a factor of 1.4 in Figure 7. As noted above, variations of this magnitude \citep{sim95,aff97} are known to exist.  Without a knowledge of the abundance,  which is usually derived for gaseous nebulae after one knows \Te\ by direct means, it is impossible to determine an accurate value of \Te\ for specific objects from low-ionization color-color diagrams.

\subsection{Systematic variations of \Te\  determined from the low-ionization color-color diagram}

Because the progression of calculated points for abundance excesses or deficits makes a loop that peaks near the predictions for the nominal abundance, abundance variations are unlikely to explain the high values of the low-ionization ratios of the lowest surface brightness WIM components. These components must have higher values of \Te. These 
high value components can be explained by photoionization processes if the illuminating stars are of higher \Tstar\ or are stars of lower temperature whose LyC radiation field mimics that of a hotter star \citep{wood04}, whereas \citet{rey99} argued that photoionization processes are insufficient.  Modifying the photoionizing radiation is possible through selective removal of photons of energies slightly greater than the ionization energy of hydrogen of 13.6 eV, a process commonly known as radiation hardening. If one introduces non-photoionization processes that come into play at these very low densities, then one doesn't need to assume a harder radiation field.

In summary we can say that the low-ionization color-color plots of Barnard's Loop, the \bub\ and the higher surface brightness components of the WIM can be explained by combinations of \Te\ and abundance variations, using available stars. In the case of the lowest surface brightness components of the WIM it is necessary to either modify the characteristic radiation field or to introduce non-photoionization heating processes. 


\subsection{Large scale line ratio variations in other galaxies}

Study of the diffuse ionized gas in other galaxies may provide some help in understanding our own WIM and the cause of the systematically higher values of \Te\ in the lowest surface brightness components.  In the study of other galaxies one has the advantage of easily looking for variations with position and these are commonly found \citep{tu00a,tu00b,ott01,ott02}, but this is at the expense of losing the diagnostically valuable tool of being able to divide the contributors into surface brightness groups and it often appears necessary to invoke non-photoionization processes to explain the observations.
The work that we report on here builds from the physics operating in a succession of photoionized objects of decreasing density and increasing scale (M~43, Barnard's Loop, the \bub, and local components of the WIM. These final conclusions can then be a jumping-off point in discussion of the diffuse ionized gas in other galaxies. 

 \section{Conclusions}

We have been able to reach several important conclusions from this study that began with new observations of the Barnard's Loop. These have provided data similar to previous studies but in a new region of the object and at higher spectrophotometric accuracy. These observations were supplemented by intensive photoionization modelling. The major conclusions are:

1. Barnard's Loop is photoionized by the most luminous stars in the Orion constellation except for  \ori\ and \oriA, whose radiation is largely absorbed locally.

2. Tests of our photoionization models on the recently well observed and intrinsically simply H~II region M~43 give a good fit to the low-ionizationization primary line ratios for stellar models close to the \Tstar\ of the complex exciting star NU Ori.

3. Barnard's Loop is similar in its properties to other regions in the Orion-Eridanus Bubble and lies at the high surface brightness end of a population of components of the WIM.

4. Our best models that explain the low-ionizationization lines fail to predict the observed strength of the [O~III] lines in both M~43 and the Barnard's Loop. This is due to a small contamination of the nebular emission by scattered light arising from M~42. 

5.The location of the Barnard's Loop observations in a low-ionization color-color diagram cannot be explained from photoionization by the most likely dominant stars unless one assumes a local Z/H enhancement of about 0.15 dex. This argues that Barnard's Loop is  enriched and possibly shaped by high velocity mass loss from evolved stars near its center.

6. The electron temperature derived from our models of Barnard's Loop and the low-ionization line ratios is about 5960$\pm$50 K, in excellent agreement with the optical/radio method result of 6100 K \citep{carl00}.

7. The population of low-ionization color-color line ratios of \bub\ and WIM components 
is enclosed by a small range of values of U, \Tstar, and Z/H.

8. Comparison of \Te\ derived by direct methods in classical high surface brightness H~II regions and a few samples of the WIM indicates that the WIM components are of comparable \Te\ to H~II regions . Only very low surface brightness H~II regions have systematically different and lower \Te . 

9. The lowest surface brightness components of the WIM in the low-ionization color-color diagram are likely to have systematically higher \Te\ than the higher surface brightness components.

10. We establish that the usual method of determining \Te\ in WIM components through I([N~II])/I(\Halpha) ratios is subject to an important uncertainty arising from known abundance variations.

11. Small changes of abundance from the reference value produce a confined short loop in the low-ionization color-color diagram for a fixed \Tstar. This means that if there is only a variation in abundance within $\pm$0.3 dex, the lowest surface brightness components of the WIM are systematically illuminated by harder radiation fields and that photoionization processes can explain the observed line ratios.

\acknowledgments
Partial financial support for GJF's work on this project was provided by National Science Foundation grants AST 0908877 and AST 0607028 and National Aeronautics and Space Administration grant 07-ATFP07-0124. CRO's work was partially supported by  STScI grant GO 10967.

{\it Facilities:} \facility{CTIO(1.5 m)}.

\appendix
\section{Recent Changes to Cloudy}
This paper makes extensive use of two newly-introduced facilities.
Cloudy can now interpolate upon large grids of stellar atmosphere spectral energy distributions.  In the case of the TLUSTY OSTAR2002 and BSTAR2006 SEDs \citep{lanz, lanz07} we can interpolate on effective temperature, surface gravity, and metallicity,  as we cannot vary abundances individually. We used TLUSTY  to be consistent with the physical calibration of the spectral classes presented by \citet{srh06}.  The interpolation methods have been generalized and many other grids of stellar SEDs are available. We have developed a domain decomposition method to compute these grids on MPI-aware parallel machines.  Each grid point is an independent model calculation and so can be done on separate computer nodes.  This results in a speedup that is of the order of the number of available processors.

The capability to compute grids of photoionization models where certain key
parameters are incremented in equidistant linear or logarithmic steps was
introduced several years ago and has been discussed in \citet{porter}.
We recently enhanced this capability by parallelizing the algorithm using the
Message Passing Interface (MPI) specification. This allows much larger grids
to be computed in parallel on distributed clusters of computers. The
calculations are set up in such a way that each core calculates a separate
model (domain decomposition) and all the results are gathered when the grid
has finished. The communication overhead is negligible since the MPI threads
only need to communicate when the grid calculation starts and finishes. Hence
this algorithm is highly efficient and scales well to high numbers of cores
for sufficiently large grids.

\section{Predicted Line Ratios}
\begin{table}
\centering
\begin{minipage}{85mm}
\caption{Predicted
Line Ratios.}
 \label{appendixB}
 \begin{tabular}{@{}ccc@{}}
 \hline
 Ion & Wavelength & I/I(\Hbeta)\\
 \hline
Ly$\alpha$ & 1216~\AA   & 12.458\\
~C~II]             &  2326~\AA  & 0.3146\\         
~Mg~II           &  2798~\AA  & 0.2749\\  
~[O~II]            &  3727~\AA  & 2.5103\\        
~[S~II]            & 4070~\AA   & 0.0389\\        
~[S ~II] & 4078~\AA &0.0130  \\          
~ \Hbeta    &  4861~\AA& 1.0000 \\           
~[N~II]  &5755~\AA &0.0109   \\         
~He~I    &5876~\AA &0.0191  \\          
~[O ~1] & 6300~\AA &0.0153   \\         
~[N~II]  &6548~\AA &0.2148  \\          
~\Halpha &    6563~\AA &2.8905  \\          
~[N~II]  &6584~\AA& 0.6337 \\           
~[S~II] & 6716~\AA& 0.3861   \\         
~[S~II]  &6731~\AA &0.2707    \\        
~[O~II]  &7323~\AA &0.0249    \\        
~[O~II]  &7332~\AA &0.0203  \\          
~ [S~III]  &9069~\AA &0.0753   \\         
~ [S~III]  &9532~\AA& 0.1869    \\        
~ He~I   &1.083~$\micron$ &0.0295    \\        
~ H~I   &1.817~$\micron$ &0.0125     \\       
~ [Ar~II] &6.980~$\micron$& 0.0500     \\       
~ [Ne~II] &12.81~$\micron$& 0.1425    \\        
~ [S ~III] &18.67~$\micron$ &0.0824   \\         
~ [S ~III] &33.47~$\micron$& 0.1916   \\         
~ [Si~II] &34.81~$\micron$& 0.0202    \\        
~ [N~II] &121.7~$\micron$& 0.0306    \\        
~ [C~II] &157.6~$\micron$ &0.3029     \\       
~ [N~II] &205.4~$\micron$ &0.0416     \\       
\hline
\end{tabular}
\end{minipage}
\end{table}

\begin{figure}
\epsscale{0.8} 
\plotone{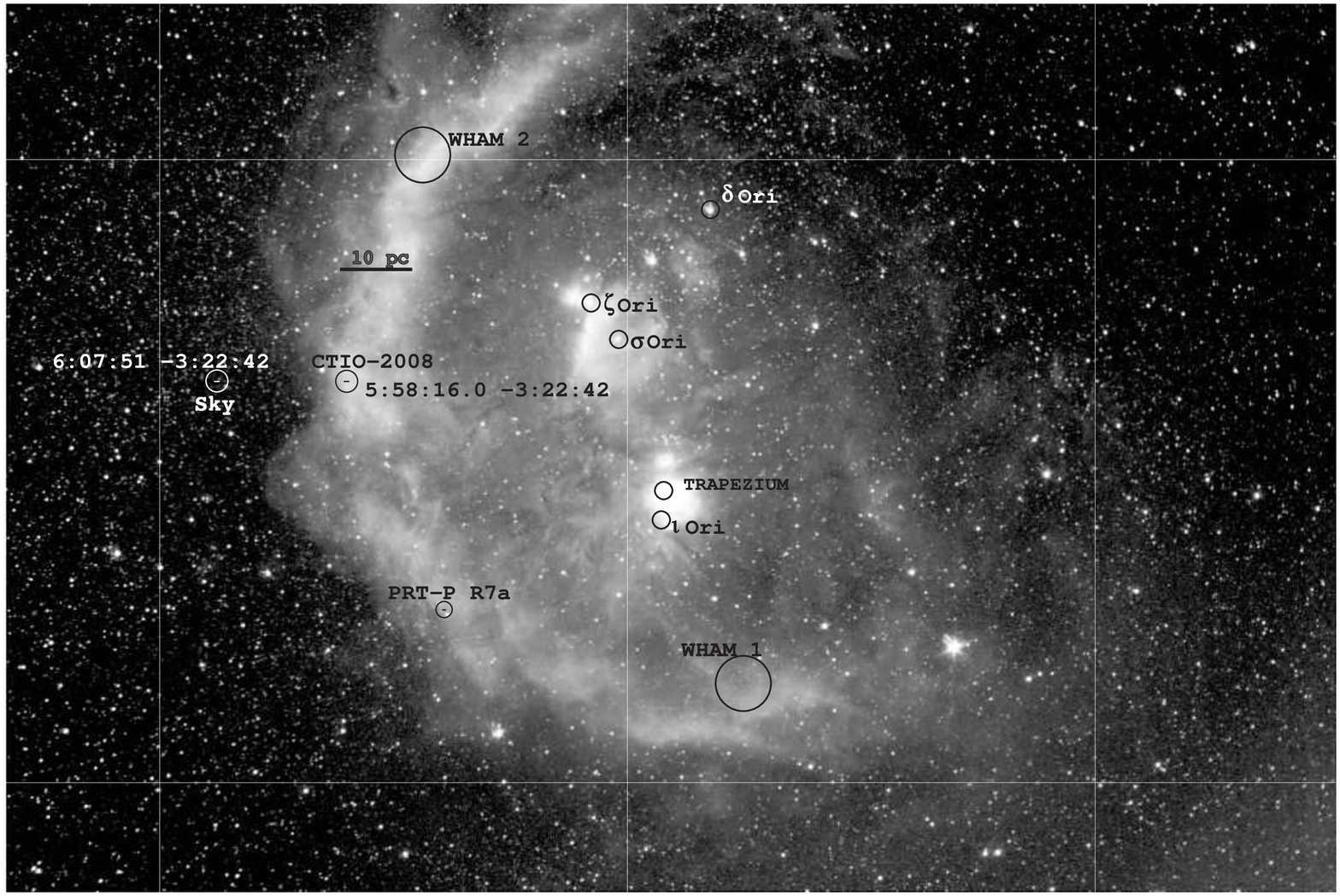}
\caption{
This 24\fdg06 x 16\fdg06 groundbased telescope image with north at the top has superimposed the positions of the slit used for our spectroscopic study(CTIO-2008) and that of \citet{pei75} (PRT-P R7a), in addition to the two WHAM apertures in the study of \citet{mad06}(WHAM 1 and WHAM2). The circle around the CTIO slit positions are there only to aid finding their locations since this long (for a spectrograph) slit is small when looking at an object as large as Barnard's Loop. The stars most luminous in Lyman Continuum (LyC) radiation are identified. The line indicating 10 pc length assumes a distance of 440 pc \citep{oh08} . The filter used for the image isolated the \Halpha +[N~II] lines and is used with the permission of Peter Erdmann (http://messier.obspm.fr/xtra/ngc/b-loop.html).}
\label{fig1}
\end{figure}

\begin{figure}
\epsscale{1.0} 
\plotone{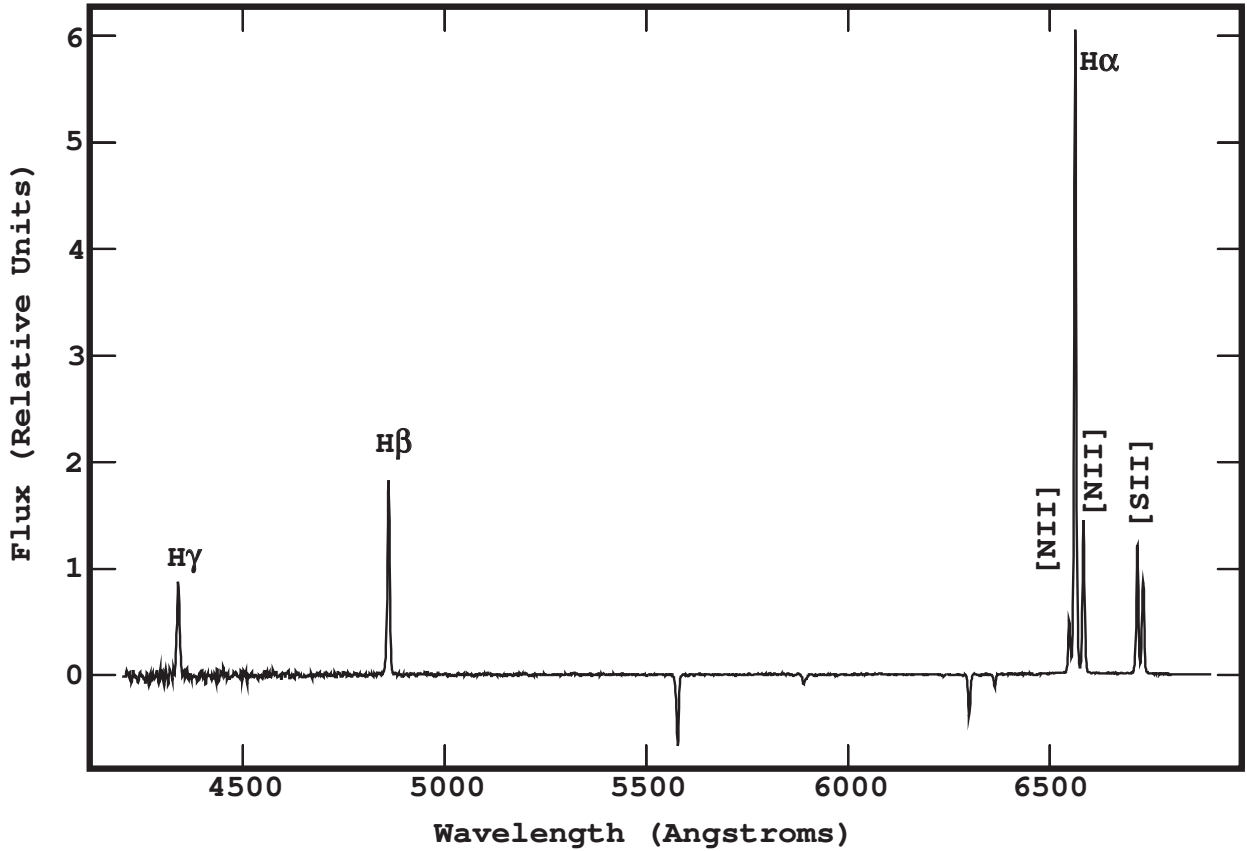}
\caption{
The final spectrum of  our new sample of Barnard's Loop is plotted in arbitrary units. Only an upper limit of about 0.007 for the ratio of flux relative to the \Hbeta\ line can be established for the [O~III] 5007 \AA. The negative signals are the result of over-subtracting the sky [O~I] lines, as discussed in the text.}
\label{fig2}
\end{figure}

\begin{figure}
\epsscale{1.0} 
\plotone{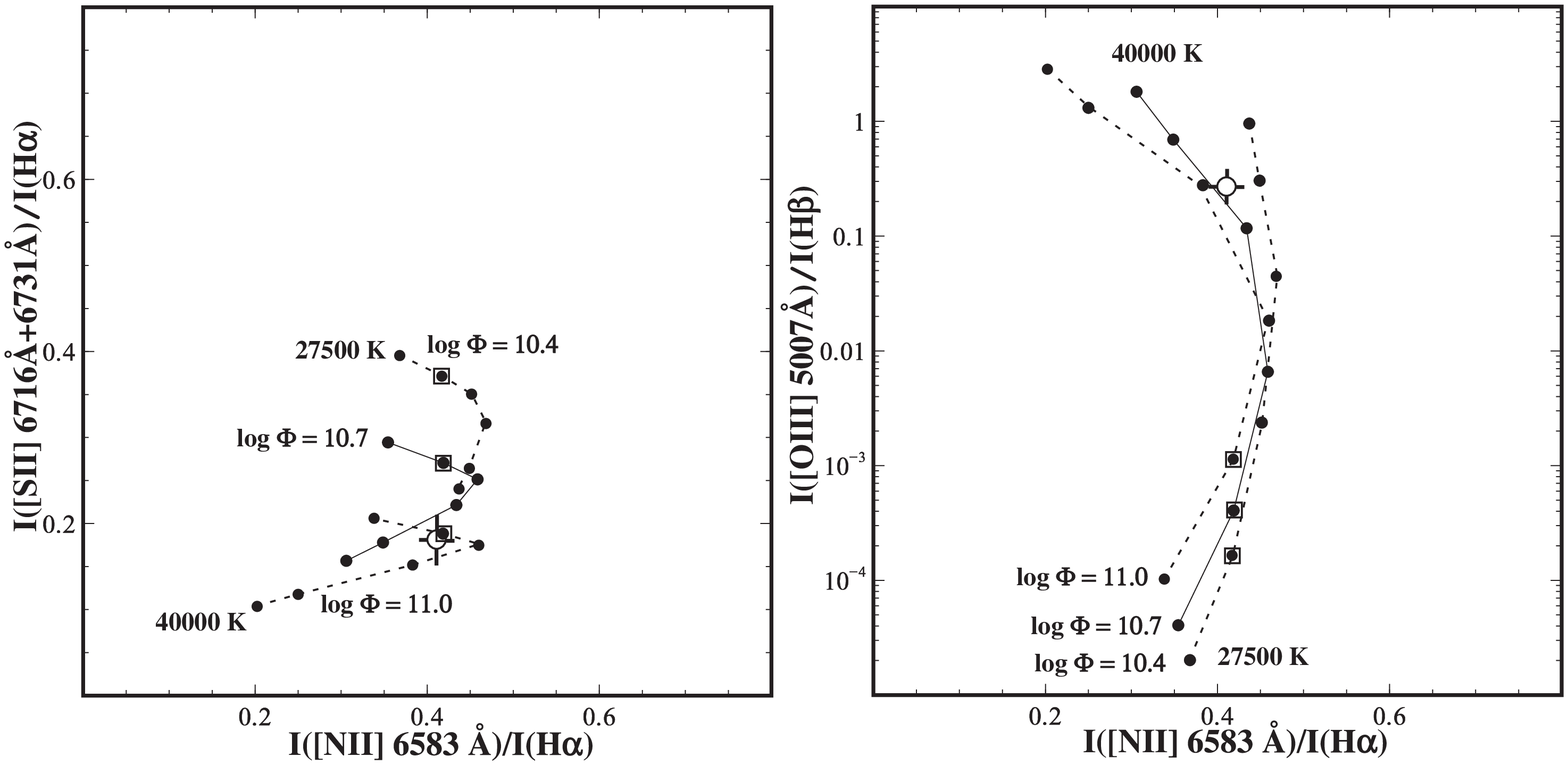}
\caption{
These two color-color plots present the results of our calculations for log $\phi$(H) = 10.4, 10.7, and 11.0 (photons cm$^{-2}$ s$^{-1}$), appropriate for M~43,  with a range of \Tstar\ of 27542 K through 40040 K  in steps of 0.0325 dex, as described in the text. The open circles with error bars give the results for the six M~43 sample spectra in \citet{od10}. The open squares surround the point calculated for \Tstar =29682 K, the closest sample to that expected from the B0.5V spectral type of NU Ori. Using the known approximate effective temperature of the star, we see that the most likely solution is log$\phi$(H) = 11.0.}
\label{fig3}
\end{figure}

\begin{figure}
\epsscale{0.8} 
\plotone{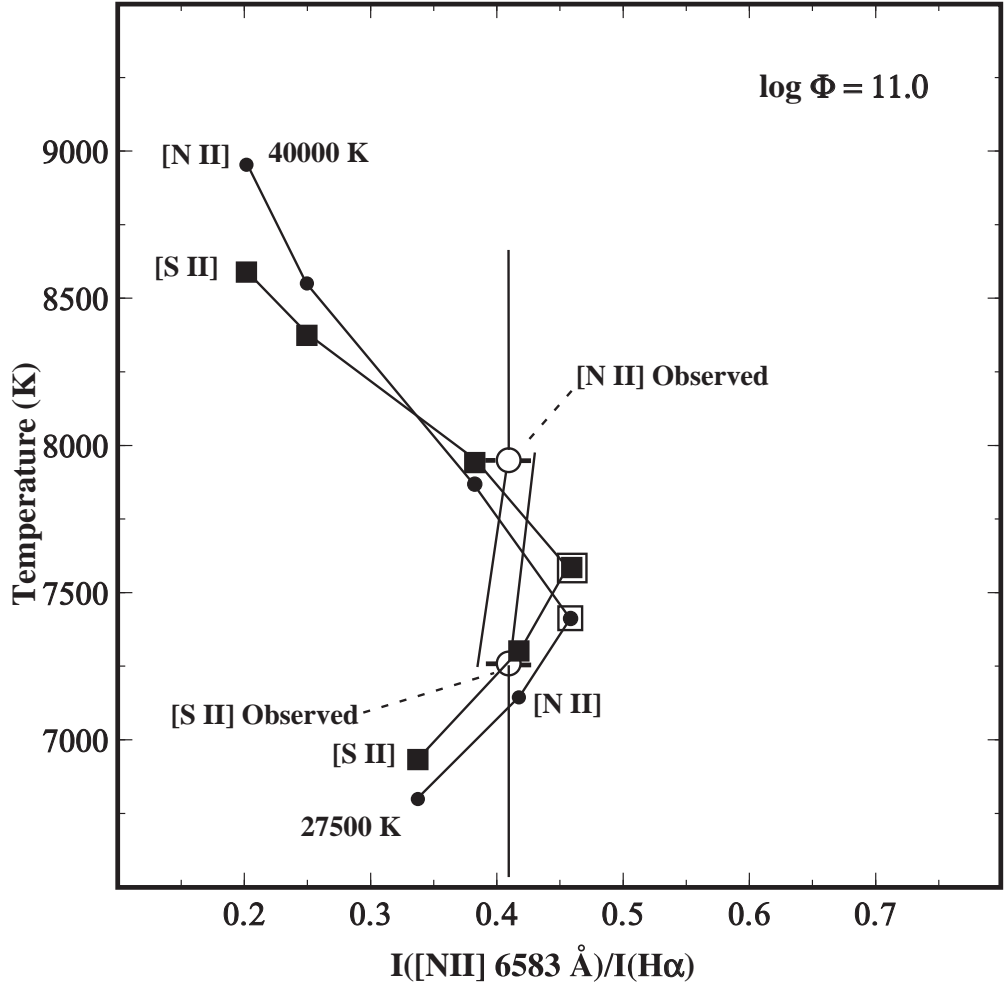}
\caption{
The M~43 predicted \Te\ for the stellar atmosphere models and the best fitting flux (log$\phi$(H) = 11.0) from Figure 3 are presented.  Filled squares are the predictions for [S~II] and the filled circles are for [N~II]. The average \Te\ values and their dispersions for M~43 as derived from the auroral to nebular line ratios of [N~II] and [S~II] by \citet{od10} are also shown.}
\label{fig4}
\end{figure}

\begin{figure}
\epsscale{0.5} 
\plotone{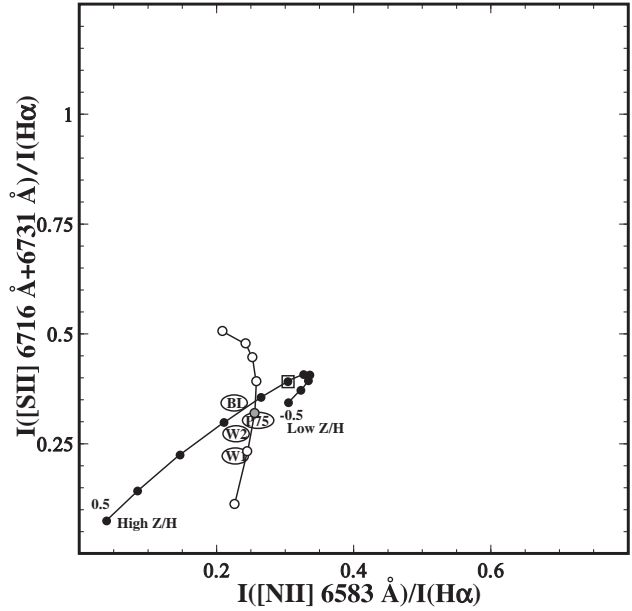}
\caption{
This low-ionization color-color plot is similar to Figure 3 (left). The series of open circles is a set of calculations using the lowest \Tstar\ in our calculations (27500) for various values of log~$\Phi$ 
( 6.7, 7.3, 7.6, 7.9, 8.2, 8.5, and 9.1 top to bottom, with the 7.9 value, favored by the model of the Barnard's Loop, shown in grey). 
The progression of filled circles are predicted line ratios for a series of heavy element to hydrogen abundance ratios (Z/H) in steps of log (Z/H)=0.1 relative to the values employed in the M~43 calculations (enclosed by an open square).  The extreme values of the log (Z/H) differences are labeled. For this sequence of calculations we have used the well defined parameters  (\Tstar = 31000 K, log~$\Phi$=7.9) for the Barnard's Loop.
We have added the observed results for this study (BL), Peimbert, et~al.'s 1975 study (P75),  and the two WHAM points (W1 and W2) lying on Barnard's Loop \citep{mad06}. In each case the uncertainty of the observed point is within the size of the enclosing ellipse.
A correction of the \citet{mad06} data by a factor of 1.67 has been made since the WHAM [S~II] data are for the 6716 \AA\ line only.
The progression from low Z/H to high Z/H is a progression towards lower \Te .
 The results shown as open circles makes it clear that no values of log~$\Phi$ with log (Z/H) = 0  matches the constraints imposed by the observed line ratios. For ease of comparison, the range of values shown are the same as in Figure 7.
}
\label{fig5}
\end{figure}

\begin{figure}
\epsscale{0.6} 
\plotone{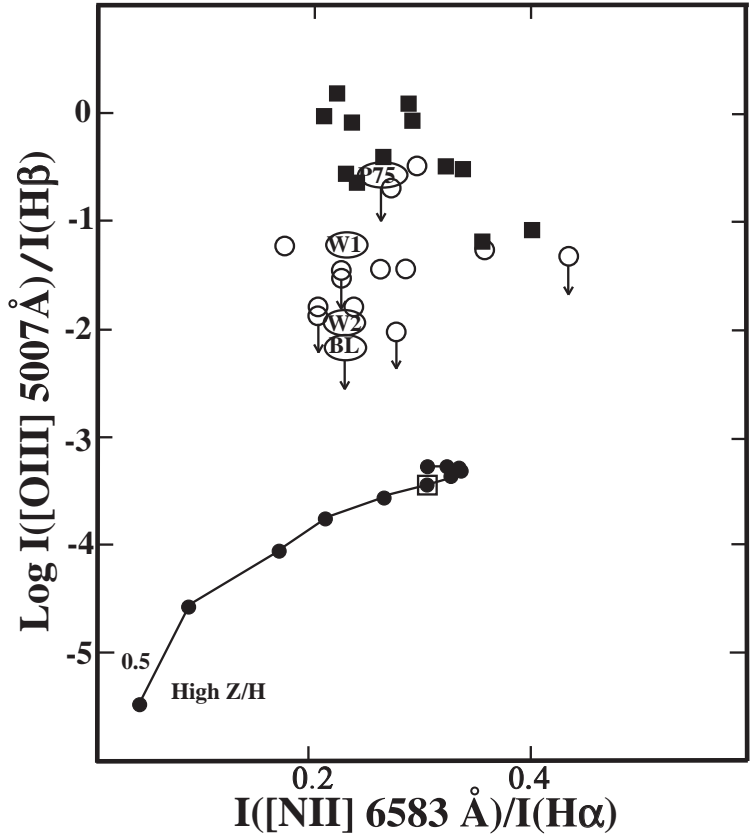}
\caption{
This high-ionization color-color plot is similar to Figure 3 (right) in the selection of the observed line ratios. 
The progression of predictions for varied values of log(Z/H)  in steps of 0.1 dex with log $\phi$(H)=7.9  (log U= -3.07) is shown as a series of filled circles with connecting lines.The largest log(Z/H) difference is labeled and the square encloses the value for the average Z/H model.  The Barnard's Loop observations are shown with the same symbols as in Figure 5. In each case the uncertainty of the observed point is within the size of the enclosing ellipse.The open circles are the \citet{mad06}
observations of regions in the \bub\ and the filled squares are the \citet{mad06} observations of low surface brightness H~II regions. In these cases the errors are less certain and probably not as great as the dispersion of the value of the samples.  Upper limits are indicated with a descending arrow. The region occupied by the low surface brightness H~II region observations generally correspond to higher values of the ionization parameter than used in the Barnard's Loop calculations.
}
\label{fig6}
\end{figure}

\begin{figure}
\epsscale{0.5} 
\plotone{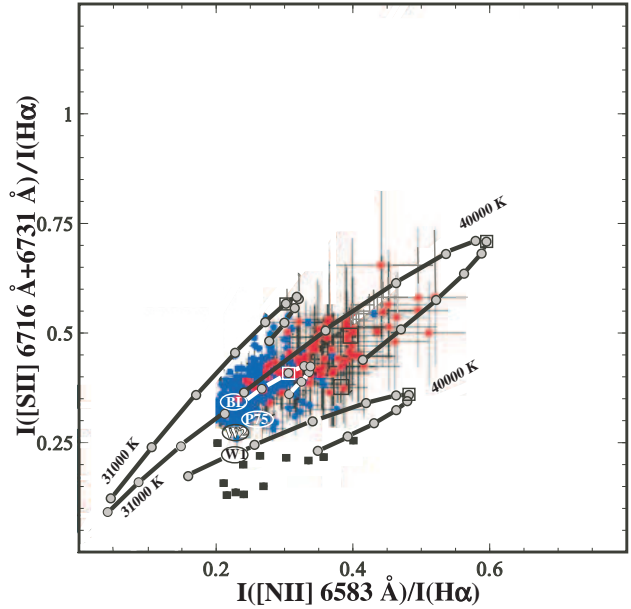}
\caption{
This figure is like Figure 5 but with additional observations of WIM components and additional theoretical models. The \citet{mad06} observations of low surface brightness H~II regions are shown as filled squares. The data from Madsen et~al.'s (2006) Figure 7 depicting ratios in the vicinity of the Orion-Eridanus Bubble are shown with error bars. The darker Madsen points to the left are higher surface brightness samples closer to the Galactic plane and falling within the Orion-Eridanus Bubble, while the lighter Madsen points are lower surface brightness samples further from the Galactic plane and are considered samples of the WIM. We also show the predictions for our photoionization models with variable abundances for stellar temperatures of 31000 K and 40000 K for two values of the ionization parameter (the upper pair of calculations for 31000 K and 40000 K are for log~U=-3.67 and the lower pair of calculations for 31000 K and 40000 K are for log~U=--3.07). The connecting lines for 31000 K and log~U=-3.07 vary in color for clarity. The pattern of variation of differences in log~(Z/H) within a sequence is the same as in Figure 5.
}
\label{fig7}
\end{figure}

\begin{figure}
\epsscale{0.7} 
\plotone{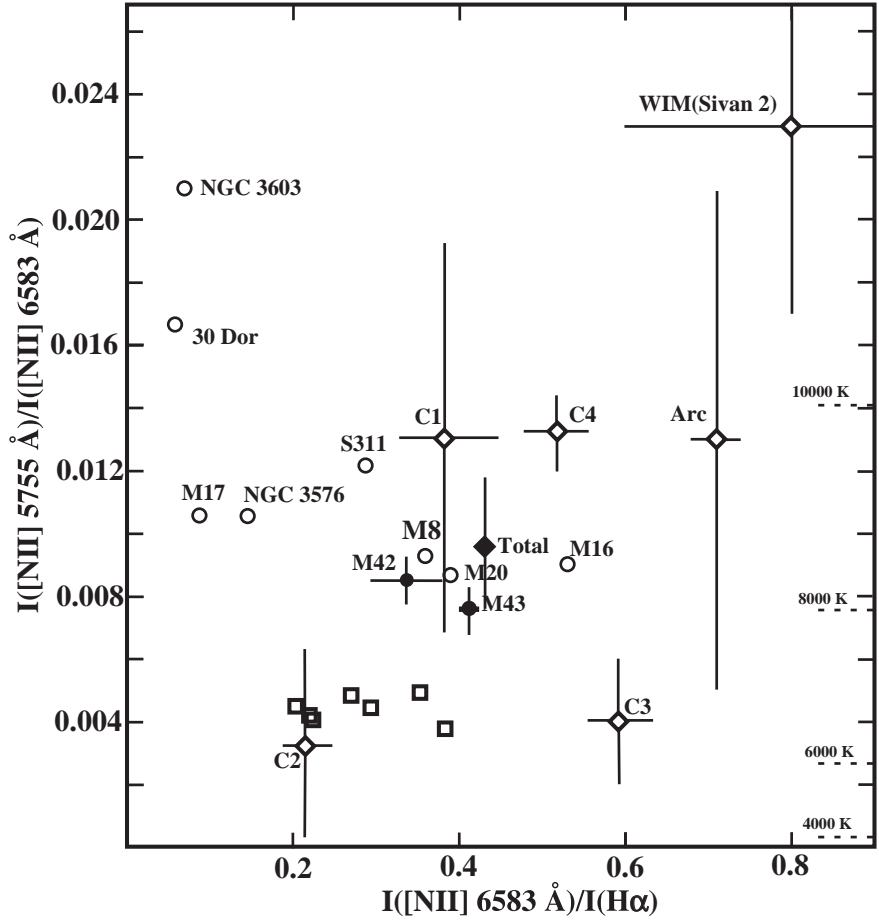}
\caption{
This figure presents the only published direct measurements of \Te\ dependent [N~II] line ratio in the WIM and compares it with the same line ratio in well-known H~II regions. Larger y values correspond to higher \Te .  Open diamonds are the individual velocity components of the WIM samples of  \citet{rey01} (C1--C4) and \citet{mad06} (Arc and WIM(Sivan 2)), the filled diamond is the result of considering all velocity components of the C1--C4 samples as a single line, and the open squares are WHAM measurements of low surface brightness H~II regions. Filled circles are for the Orion objects M~42 and M~43 \citep{od10}, and open circles are from the studies of \citet{gar04,gar05,gar06a,gar06b}. There is no indication that the WIM components are systematically higher \Te\ than the higher density well-known H~II regions. The dashed lines on the right indicate the  I([N~II] 5755 \AA)/I([N~II] 6583 \AA) ratios expected at various \Te .
}
\label{fig8}
\end{figure}

\end{document}